\documentclass[conference]{IEEEtran}
\IEEEoverridecommandlockouts
\usepackage{cite}
\usepackage{amsmath,amssymb,amsfonts}
\usepackage{algorithmic}
\usepackage{graphicx}
\usepackage{textcomp}
\usepackage{xcolor}

\usepackage{bm}
\usepackage{amsmath}
\usepackage{amsfonts}
\usepackage{amssymb}

\def\BibTeX{{\rm B\kern-.05em{\sc i\kern-.025em b}\kern-.08em
    T\kern-.1667em\lower.7ex\hbox{E}\kern-.125emX}}
\begin{document}

\title{Precheck Sequence Based False Base Station Detection During Handover: A Physical Layer Security Scheme\\
}

\author{
\IEEEauthorblockN{Xiangyu Li\IEEEauthorrefmark{1}\IEEEauthorrefmark{2}, Kaiwen Zheng\IEEEauthorrefmark{1}, Sidong Guo\IEEEauthorrefmark{1}, Xiaoli Ma\IEEEauthorrefmark{1}}
    \IEEEauthorblockA{
    \IEEEauthorrefmark{1}School of Electrical and Computer Engineering, Georgia Institute of Technology, Atlanta, USA\\
    \IEEEauthorrefmark{2}Georgia Tech Shenzhen Institute, Tianjin University, Shenzhen, China\\
    e-mail: \{xli985, kzheng71, sguo93, xiaoli\}@gatech.edu
    }
}


\maketitle
\begin{abstract}
False Base Station (FBS) attack has been a severe security problem for the cellular network since 2G era. During handover, the user equipment (UE) periodically receives state information from surrounding base stations (BSs) and uploads it to the source BS. The source BS compares the uploaded signal power and shifts UE to another BS that can provide the strongest signal. An FBS can transmit signal with the proper power and attract UE to connect to it. In this paper, based on the 3GPP standard, a Precheck Sequence-based Detection (PSD) Scheme is proposed to secure the transition of legal base station (LBS) for UE. This scheme first analyzes the structure of received signals in blocks and symbols. Several additional symbols are added to the current signal sequence for verification. By designing a long table of symbol sequence, every UE which needs handover will be allocated a specific sequence from this table. The simulation results show that the performance of this PSD Scheme is better than that of any existing ones, even when a specific transmit power is designed for FBS.

\end{abstract}


\section{Introduction}
Globally, the fifth Generation Mobile Communication System (5G) is providing greener networks \cite{alsharif2014} with increasingly high quality of service (QoS), mainly in terms of higher throughput, spectral efficiency and energy efficiency, and lower complexity in signal transmitting and processing \cite{li2016}. However, due to backward compatibility, 5G still inherits many mechanisms from previous generations and this is where some security problems may arise. One of the most important mechanisms is the reselection of cell \cite{Udoh2018} or base station (BS) for user equipment (UE). The basic principle lies in the nature of UE connection - to find better QoS from surrounding environment. This mechanism is pervasive for devices around and can be of vital significance for the duration of connectivity. However, the weaknesses in this mechanism also create opportunities for potential attacks - the false base station (FBS) attack.


FBS, also referred to as pseudo base station (PBS) or malicious base station (MBS), is an illegal BS which aims at attacking surrounding or targeted devices passively or actively over radio access networks (RANs). It has the ability to utilize the potential weaknesses in the network structure to force UE's connection with itself instead of the legal base station (LBS). Besides, it is difficult to predict when, where and how the threats from FBS will appear.


\section{Background}

\subsection{Handover Mechanism}
\begin{figure}[ht]
    \centering
	\includegraphics[width=0.4\textwidth]{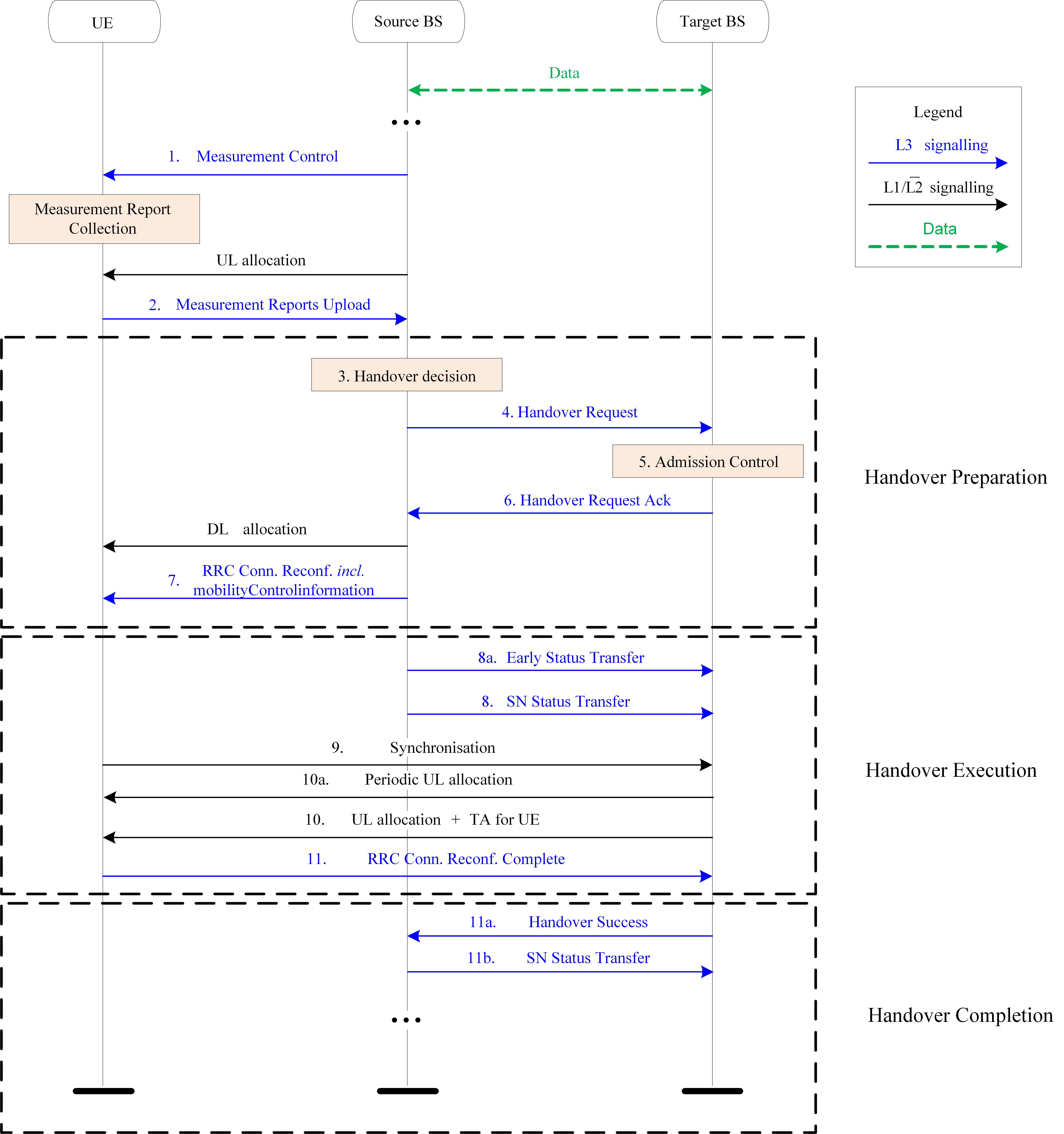}
	\caption{Handover procedure}
	\label{fig:1_1 Handover Procedures}
\end{figure}

Handover is a process in communications where a transition is made to shift the connection from the current cell to another cell without ending session. In order to secure this process, the 3rd Generation Partnership Project (3GPP) formulates a series of data, e.g. Measurement Report (MR), to help determine whether a transition should be executed. Three stages, Handover Preparation, Handover Execution and Handover Completion in Figure \ref{fig:1_1 Handover Procedures} summarize how UE is shifted from source BS to target BS according to \cite{3gpp.36.300}. The 3GPP handover standard also includes Mobility Management Entity (MME) and Serving Gateway (SG) in the complete Technical Specification. As the main procedures of FBS attack occur before MME and SG get involved, we only focus on the interactions among UE, source BS and target BS.

Unfortunately, the source of system information is never authenticated even in nowadays 5G networks, which is a weakness in the handover mechanism. The backward compatibility has not been properly dealt with yet, FBS may intercept the system information of a nearby LBS and replace it in a similar or higher signal power, so that one or more UEs will be illegally connected. In order to avoid such FBS attacks brought by authentication failure in network or link layer, extra verification schemes in physical layer need to be introduced for more accurate FBS detection.




There are many types of attacks that FBS can initiate, including Impersonation, Intercept and Eavesdropping \cite{Abodunrin2015} etc. While all devices in the network, e.g. BS, Relay BS, or even the core network, are exposed to such attacks, the majority victims are mobile devices. FBS has the ability to force a UE's connection from an LBS to itself. Basically, this is realized by filching the broadcast messages of a selected LBS and increasing the transmit power to make UEs choose FBS. 



\subsection{Related Work}
Some cryptographic detection schemes are introduced in 3GPP 5G Specification \cite{3gpp.33.512} to secure handover. However, these cryptography-based models may either increase the complexity or fail to defend against the updated attacks in real-time and multi-cast cases. The previous detection schemes fall into three categories \cite{nakarmi2021} as follow:
	
(1) \textbf{UE-based detection scheme}. Many previous detection schemes were in this category since they relied on the data at UE side.
In \cite{Zhuang2018}, a pragmatic radio frequency (RF) fingerprinting-based FBS detection approach was investigated and improvements were made in \cite{Ali2019enabling} to enable carrier frequency offset (CFO)-based schemes in time-critical scenarios. However, these schemes could be time-consuming and computationally-inefficient.


(2) \textbf{Crowd-Sourced scheme}. With data collected from distributed UEs, the processing of data is done by the source BS or other  data centers. A network of stationary measurement units and an application for mobile phones were studied in \cite{Dabrowski2014} to detect IMSI catcher. What in common was that both implementations required scanned data feedback to a central processing unit. A machine learning-based IMSI catcher detection system based on publicly available data was presented in \cite{Do2015}, which combined three detectors – Off-line-learning detector, Anomaly detector and Ensemble detector.

(3) \textbf{Network-based scheme}. Information from both UEs and BSs is processed in the core network or cloud servers instead of the local BS. 
A cloud-server-based detection method was provided in \cite{Al2015}, which required BSs and UEs to transmit information to the cloud server. This uploading of information from both UEs and BSs increased uplink load, especially when there were no FBS attacks.

Only a few researches focused on the security of handover process where FBS attacks may occur. A mathematical model was developed in \cite{Saedi2020} for FBS attack where several LBSs supported the network connection in the presence of vehicles. In \cite{saedi2022}, the positions of LBSs and FBS random were made random and similar received signal strength (RSS) of a platoon of vehicles was generated. However, no detection schemes have been proposed, which has left room for future researches.

\textit{Notation:} Boldface, lower-case letters denote column vectors and boldface upper-case letters denote matrices. The superscript $(\cdot)^{T}$ represent the transpose.

\section{System Model} \label{system model}

\subsection{Device Deployment}
\begin{figure}[ht]
    \centering
	\includegraphics[width=0.4\textwidth]{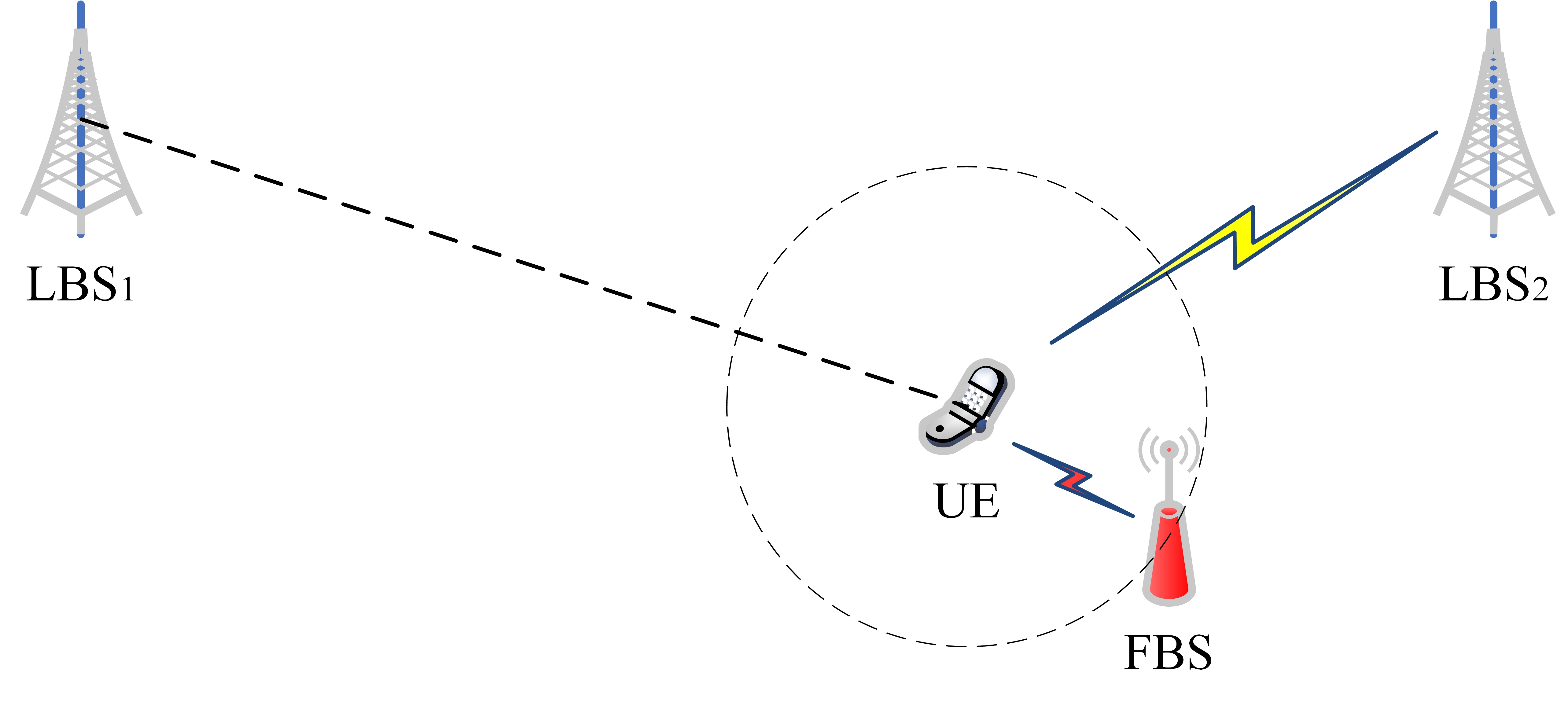}
	\caption{The deployment of LBSs, FBS and UE}
	\label{fig:2_1 System Model}
\end{figure}

We consider a system that is composed of a source BS, a target BS, an FBS and a UE, as shown in Figure \ref{fig:2_1 System Model}. The source BS, i.e. LBS$_{1}$, and the target BS, i.e. LBS$_{2}$, are respectively located in the center of their own cell. UE is at the junction area of these two adjacent cells and FBS is situated randomly in an annular region from a distance to UE.

Initially, while UE is connected to LBS$_{1}$, it also detects stronger signals coming from a neighboring LBS$_{2}$. FBS, which is closer to UE, wiretaps parameters of LBS$_{2}$ and disguises its own signal parameters as those from LBS$_{2}$. Because of this impersonation, UE detects two signals that originate from the same LBS$_{2}$ and it is the nature of UE to be connected via a stronger signal. In this way, UE may be connected to FBS.

\subsection{Serial Transmission Model}

\begin{figure}[ht]
    \centering
	\includegraphics[width=0.45\textwidth]{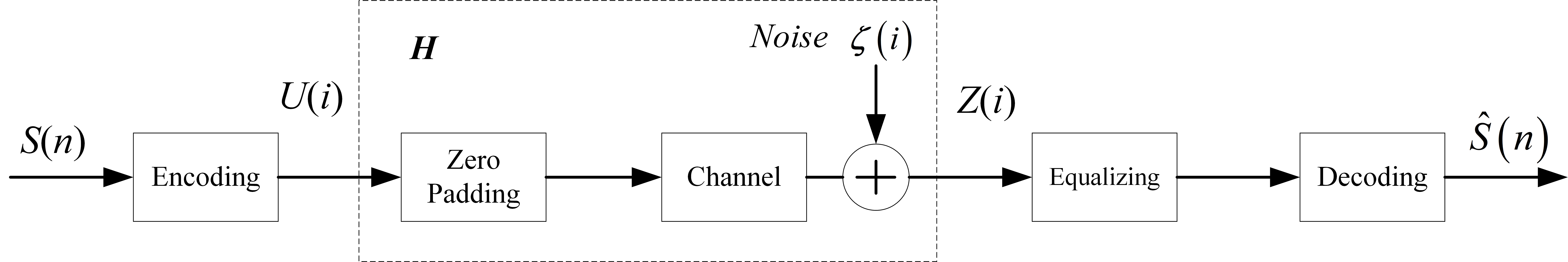}
	\caption{Serial transmission model}
	\label{fig:2_2 Zero Padded Block Transmission System}
\end{figure}

We follow an improved single-carrier serial transmission model in \cite{wang2004}. The index $n$ is used for symbol streams of possible different rates. Initially, a serial information symbols stream $S(n)$ of goes through an error-control encoder, whose output is defined as $U(n)$. If it is not encoded, $U(n)=S(n)$. In each block, define the channel order as $L_{c}$, block size as $N_b$, and the total length of linear convolution is $P = N_b + L_{c}$. The sequence $U(n)$ is grouped into these blocks
\begin{equation}
    \bm{U}(i) = \left[ U(iN_b),U(iN_b+1),...,U(iN_b+N_b-1) \right]^T.
\end{equation}
The $i$-th observed block can be listed as
\begin{equation}
    \bm{Z}(i) = \bm{H}\bm{U}(i) + \bm{\zeta}(i)
\end{equation}
where each entry of the Toeplitz convolution matrix $\bm{H}$ is $\left[ \bm{H} \right]_{p,n} = H(p-n)$, and the AWGN is $\bm{\zeta}(i) = \left[ \zeta(iP),\zeta(iP+1),...,\zeta(iP+P-1) \right]^T$. The structure of this model is depicted in Figure \ref{fig:2_2 Zero Padded Block Transmission System}.

\subsection{Assumptions for FBS}
\paragraph{Assumption 1} FBS initiates attacks by imitating a neighboring LBS, i.e. LBS$_{2}$ of the source BS, i.e. LBS$_{1}$. 
\begin{itemize}
\item When FBS is imitating LBS$_{2}$, it will focus only on LBS$_{2}$ and not wiretap or imitate transceiving messages of other possible LBS - LBS$_{1}$, LBS$_{3}$ etc. This is because FBS is relatively closer to LBS$_{2}$; otherwise, there will be too much workload for FBS.
\end{itemize}

\paragraph{Assumption 2} In order to attack as many UEs as possible, FBS wiretaps UEs’ direct transceiving synchronization messages and UL allocation messages with LBS$_{2}$.
\begin{itemize}
\item UEs’ messages sent from and to other LBSs, i.e. LBS$_{1}$, and possible LBS$_{3}$ etc. will not be wiretapped by FBS.

\item FBS begins sending UL allocation messages and time advances (TA) messages to a UE only after it learns that the UE is sending synchronization messages to the LBS$_{2}$. FBS does not send these messages after it has wiretapped LBS$_{2}$'s complete UL allocation messages.

\item Due to broadcasting, UL allocation messages that FBS sends to a certain UE can be detected by other UEs only as MR messages from FBS, but will not be responded.
\end{itemize}

To compare the performances of different detection schemes and for simplicity, we focus on the scenario where there is only one UE. The extension to multiple UEs is left as future work.

\section{Proposed Detection Scheme}

\subsection{Fundamental Component}

\begin{figure}[ht]
    \centering
	\includegraphics[width=0.45\textwidth]{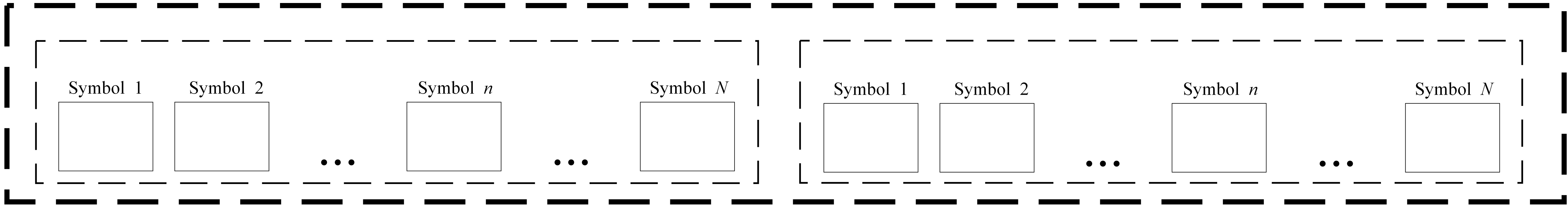}
	\caption{Table of symbols with repeated patterns}
	\label{fig:4_2_1 Symbols in Table of Sequence}
\end{figure}

Based on the handover in 3GPP Specification \cite{3gpp.36.300}, we design a table of sequence which consists of a series of symbols to verify the legitimacy of received signals. To ensure symbols are selected in a successive order, the table of sequence consists of two identical parts of symbols. The first half of sequence from $Symbol$ $1$ to $Symbol$ $N$ on the left is the same as the second half on the right, as shown in Figure \ref{fig:4_2_1 Symbols in Table of Sequence}.

\begin{figure}[ht]
    \centering
	\includegraphics[width=0.45\textwidth]{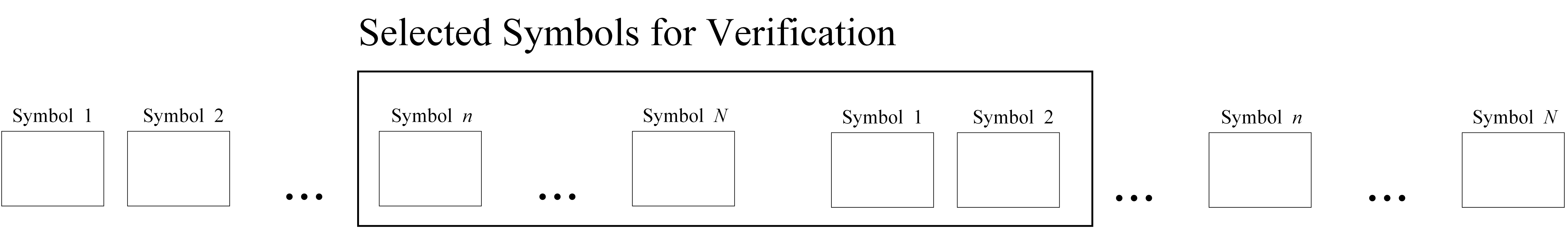}
	\caption{Selected symbols for verification}
	\label{fig:4_2_2 Selected Symbols for Verification}
\end{figure}

As shown in Figure \ref{fig:4_2_2 Selected Symbols for Verification}, before LBS sends signals, a fixed number of symbols, referred to as \textbf{\textit{selected symbols}}, are first chosen from the table of sequence. These symbols can be seen as \textbf{\textit{precheck sequence}} ahead of regular information in signals. The beginning symbol can be chosen from the first half of sequence, and ends within the second half. By leveraging $M$-QAM modulation, every $\sqrt{M}$ bits will be modulated and demodulated into each symbol.

\subsection{Adversary Attack Description}

FBS imitates the target BS by overhearing the contents that were and are being transmitted by the target BS, and replaying them. We assume FBS knows that some symbols are attached ahead of the regular information for signal verification. Not only the regular information such as cell ID, TA, UL Grant etc., but also the selected symbols will be wiretapped by FBS in order to generate similar or even almost accurate sequence as that sent from the target BS. However, reasonable time delay is considered by UE for security. UE can estimate the anticipated arrival time of the signal, based on the distance between itself and the target BS. After UE sends the synchronization message to the target BS, the target BS will respond to UE with periodic UL allocation messages. Thus, if FBS overhears the target BS before transmitting the exact symbols, UE can find the longer time delay of a illegal signal.

In order to let UE suppose the signals from FBS are legal, as soon as it detects synchronization information sent from UE, FBS will choose continuous symbols from the known table of sequence and transmit them to UE together with the regular information. This will let the signal from FBS be received by UE during anticipated time. In this way, FBS may succeed disguising itself as the target BS without being discovered.

\subsection{Improved Handover Scheme with Sequence Verification}

In this scheme, these received symbols are added ahead of regular information. Every time it selects symbols, the target BS will have the selected sequence of symbols randomly begin with different symbols in the table. As shown in Figure \ref{fig:4_2_2 Selected Symbols for Verification}, the selected symbols begin with Symbol $n$ and end with Symbol $2$. This will make it difficult for FBS to predict what the transmitted precheck sequence exactly is; however, the number of selected symbols is fixed and is known to FBS.

\begin{figure}[ht]
    \centering
	\includegraphics[width=0.4\textwidth]{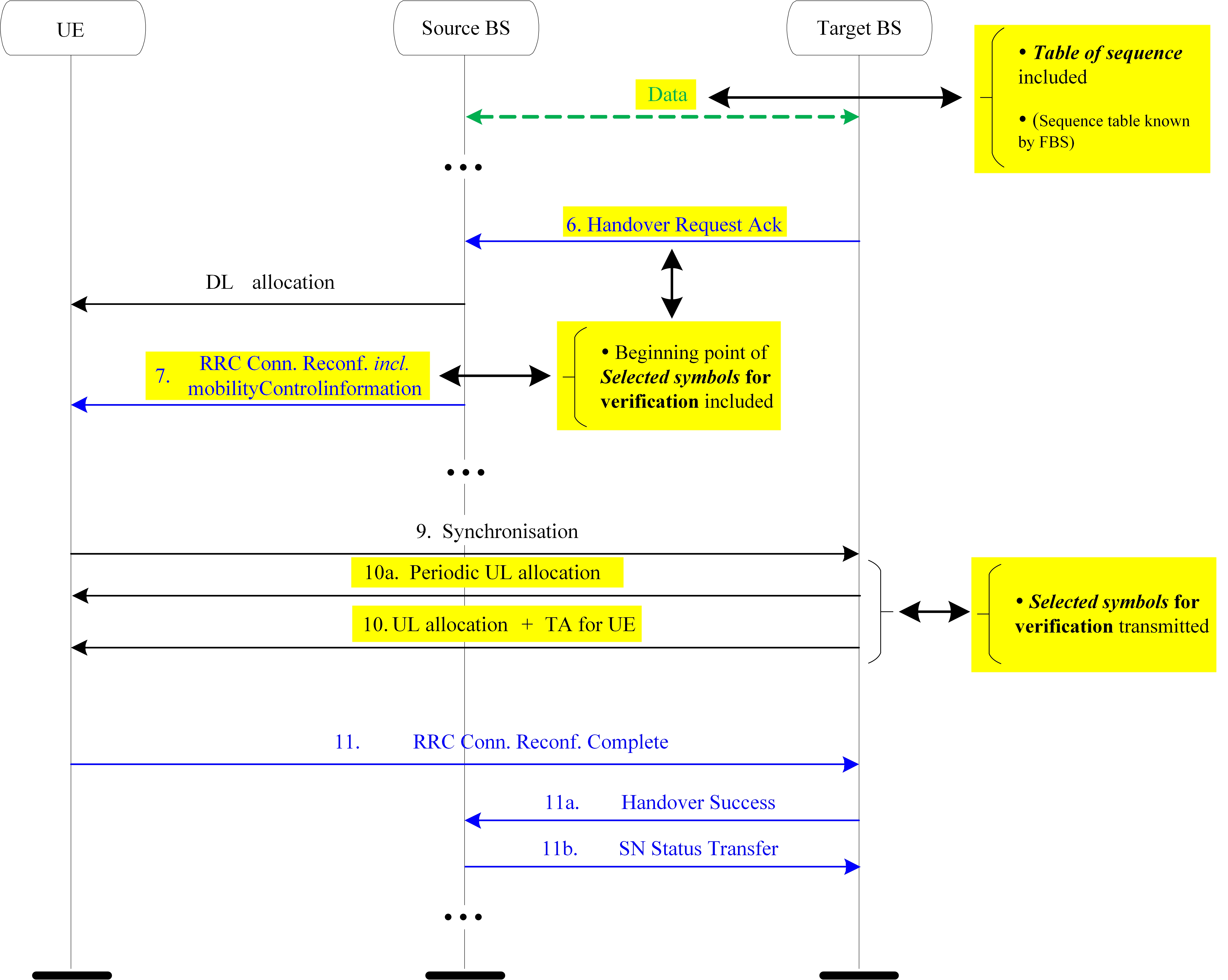}
	\caption{Handover procedure with sequence verification}
	\label{fig:4_2_3 Handover procedures with sequence verification}
\end{figure}

The core of this detection scheme lies in the consistency of anticipated received symbols and actual received symbols. Key steps are depicted in Figure \ref{fig:4_2_3 Handover procedures with sequence verification}: 
(1) Before handover, the target BS shares the whole table of symbols so that it is known by any devices including FBS. 
(2) In \textit{Handover Request Ack} transmitted from the target BS to the source BS, the beginning symbol index of selected symbols and the total number of selected symbols are included. 
(3) The source BS determines selected symbols and transmits them together with the regular information back to UE. The selected symbols transmitted by the source BS are considered as \textbf{\textit{standard precheck sequence of symbols}} for later verification. 
(4) Upon receiving the signal from source BS, UE transmits synchronization information to target BS and waits for a response. (5) The target BS transmits selected symbols to UE with UL allocation or TA information.

During verification, UE will check the received signal in the form of symbols and compare them with the previously-received sequence of symbols from the source BS. Based on the distance between UE and the target BS, UE can estimate the expected arrival time period of the received signal. On the contrary, if a signal arrives at UE much earlier or later than the expected time, UE can notice it and then mark it as an illegal signal. With verification of both received bits and arrival time, the source of signal can be more easily authenticated. For this Precheck Sequence-based Detection (PSD) Scheme comes a third assumption:

\paragraph{Assumption 3} Considering generating cost and computing complexity when transmitting sequence from target BS to source BS, target BS will not generate a complete and new table of symbols for every UE. Instead, the general sequence table is unchanged, kept in storage and public to any devices.
\begin{itemize}
\item Though FBS can know the whole table of sequence, it does not know what the beginning symbol is.

\item Every time the source BS decides to hand off UE to the target BS, the target BS only needs to inform the source BS the beginning symbol index before transmitting selected symbols and other messages.

\item When FBS imitates the target BS, it only overhears and focuses on the communication between the target BS and UEs. As FBS is closer to the target BS than other LBSs, it does not know about the transmission of other LBSs.

\end{itemize}

\section{Numerical Results and Discussion}
The evaluation criteria of legal sequence of symbols involves bit error rate (BER). Especially for the case where two signals, one from the target BS and the other from FBS, arrive at nearly the same time, UE compares two sets of selected symbols with the aforementioned \textbf{\textit{standard precheck sequence of symbols}} respectively. The signal with higher BER is regarded as illegal and coming from FBS.

\subsection{Results Analysis and Comparison}
Let the table of symbols generated by gray mapping, and modulated via 16-Quadrature Amplitude Modulation (16-QAM). We define the length of table of symbols as 32, the block size $N=4$ and the channel order $L_{ZP}=2$. Under 16-QAM, a total of 2 blocks, i.e. 8 symbols, are added ahead of regular transmitted information. Each simulation below uses MATLAB R2022a and is conducted for 10,000 realizations.

\begin{figure}[ht]
    \centering
	\includegraphics[width=0.4\textwidth]{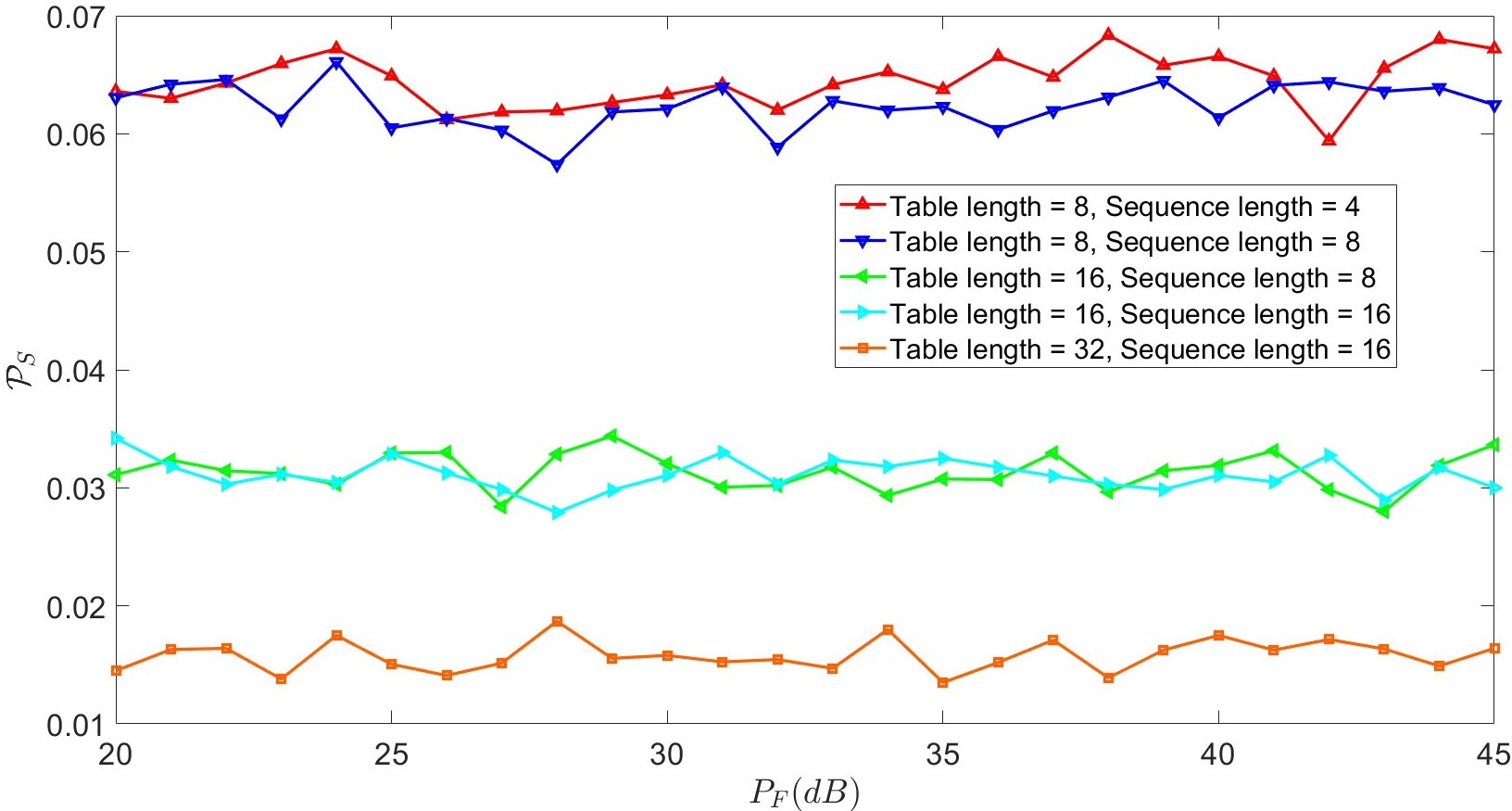}
	\caption{PSD Scheme under different table lengths and sequence lengths}
	\label{fig:4_4_1 Sequence scheme under different table length and sequence length}
\end{figure}

Figure \ref{fig:4_4_1 Sequence scheme under different table length and sequence length} depicts the successful cheating rate (SCR) of the PSD Scheme under different sets of table length and sequence length. In a stable communication system with comparatively low BER, it can be inferred that the change of sequence length does not affect detection performance much when the table length stays the same. However, with fixed sequence length, the increase of table length leads to lower SCR thus better detection performance.

The PSD Scheme is compared with representative schemes which have been proposed in \cite{li2017},\cite{pradeepkumar2017} and \cite{huang2018} respectively. For the above references, since the most vulnerable occasions differ due to the transmit power of FBS, we make comparisons separately in the figures below.

\begin{figure}[ht]
    \centering
	\includegraphics[width=0.4\textwidth]{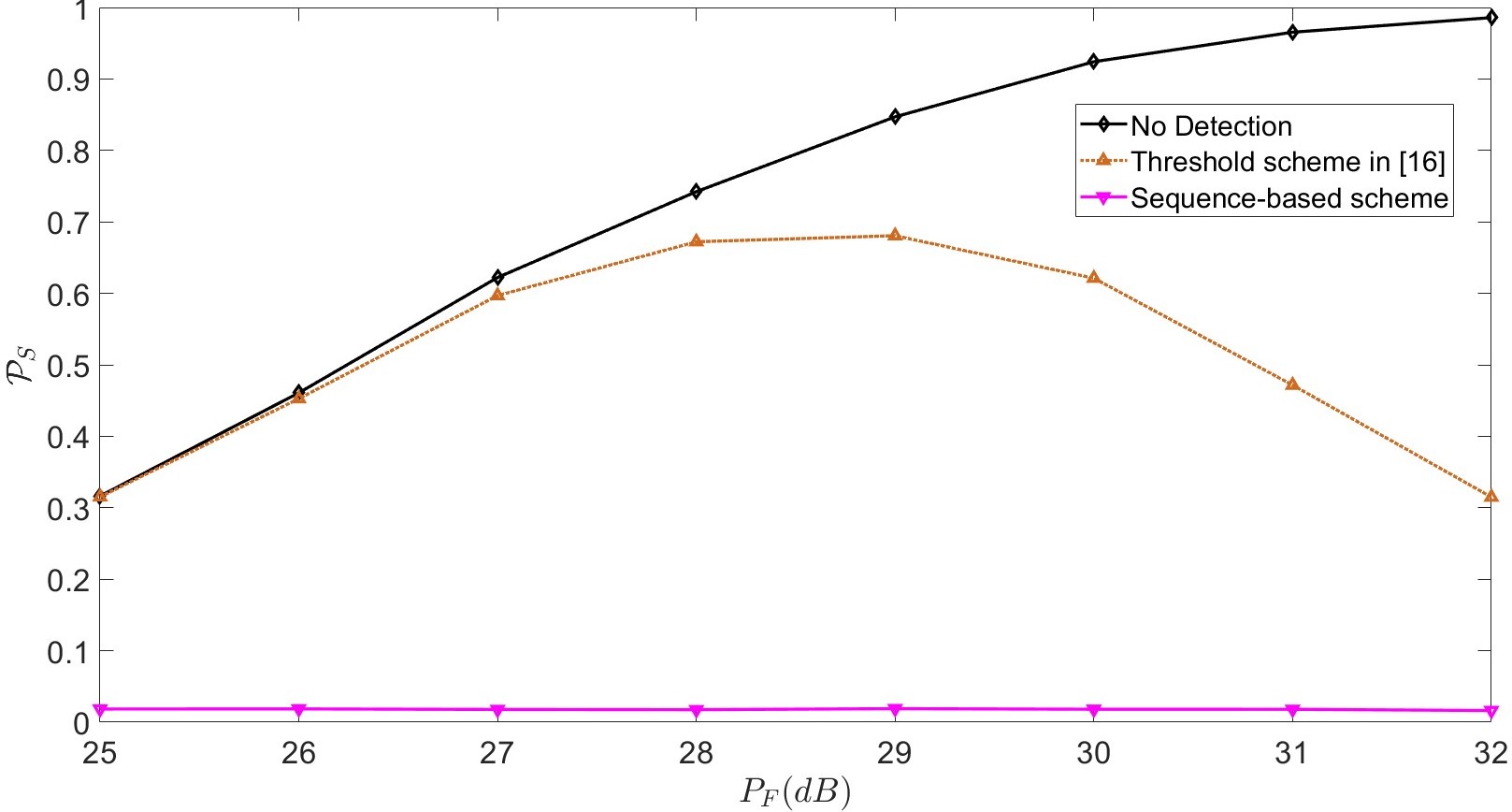}
	\caption{PSD Scheme compared with [16]}
	\label{fig:4_4_2 Sequence Scheme compared with [41]}
\end{figure}

First, comparisons are made between the PSD Scheme and the scheme in \cite{li2017}. The scheme in \cite{li2017} assumes that FBS messages usually have the signal strength which is higher than a certain value, which is approximately three standard deviation above the average. As shown in Figure \ref{fig:4_4_2 Sequence Scheme compared with [41]}, if FBS appropriately calculates its transmit power and makes UE's RSS within the legal range, UE can still be easily attacked. However, with PSD Scheme, incorrect sets of sequence symbols can be directly judged as illegal. Therefore, the SCR under this scheme will not increase with the transmit power of FBS.

\begin{figure}[ht]
    \centering
	\includegraphics[width=0.4\textwidth]{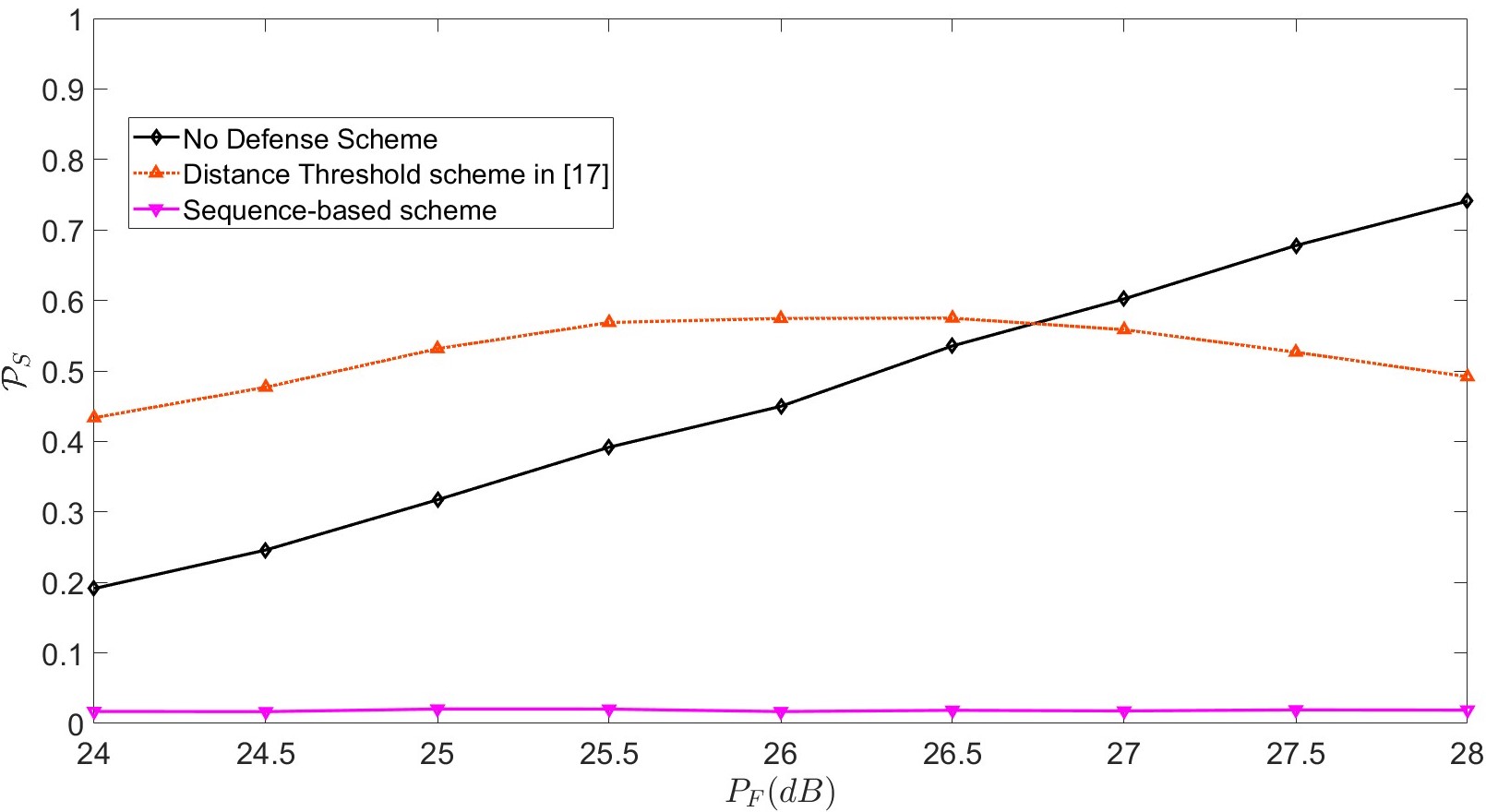}
	\caption{PSD Scheme compared with [17]}
	\label{fig:4_4_3 Sequence Scheme compared with [42] and scheme with LF}
\end{figure}

\begin{figure}[ht]
    \centering
	\includegraphics[width=0.4\textwidth]{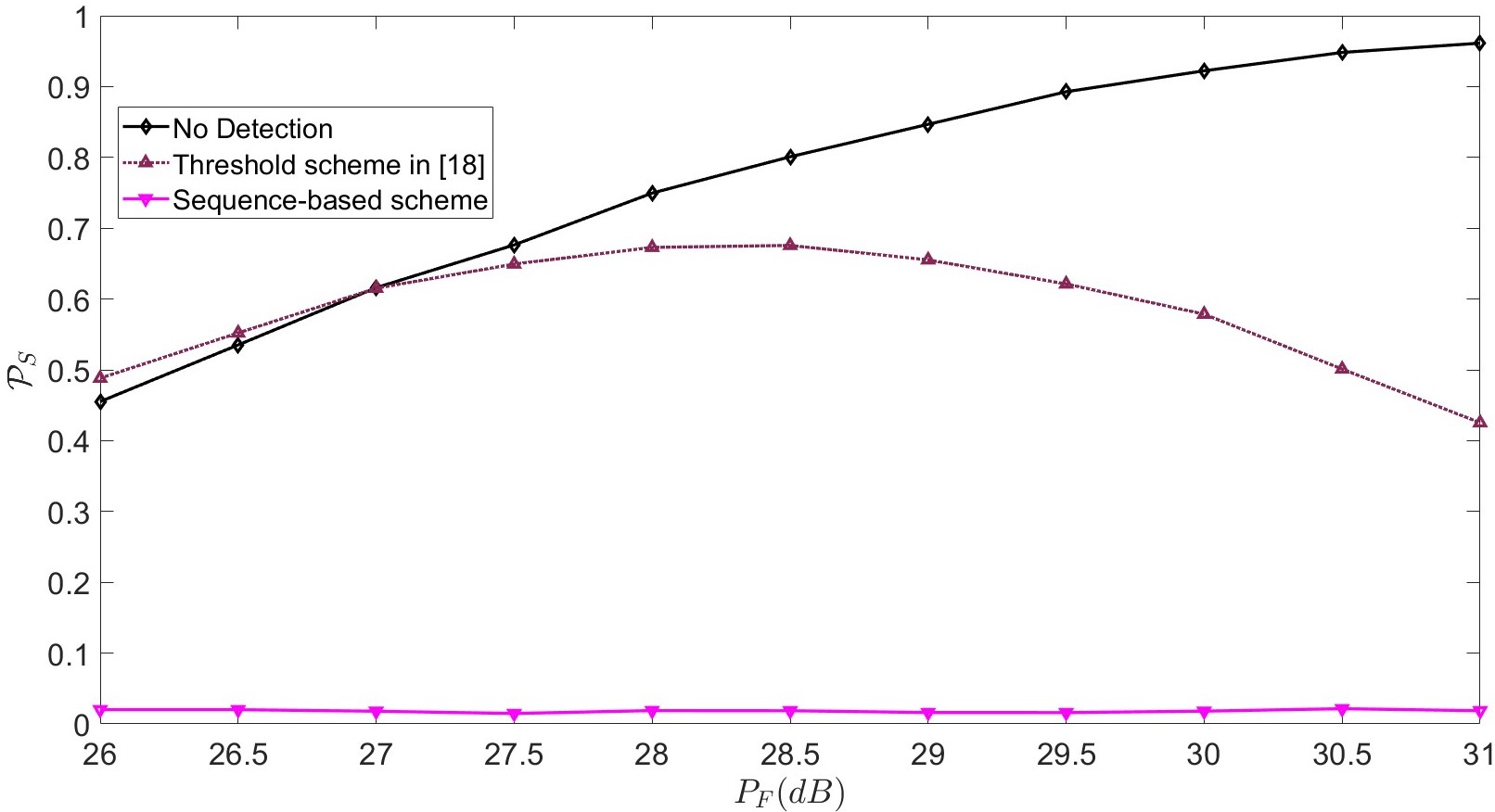}
	\caption{PSD Scheme compared with [18]}
	\label{fig:4_4_4 Sequence Scheme compared with [43]}
\end{figure}

Next, a distance threshold-based scheme has been proposed in \cite{pradeepkumar2017}, which uses transmission model to calculate the distance of UE from the target BS and sets a distance threshold. Besides, a suspicious region has been designed in \cite{huang2018} with a small pre-set value similar to the false alarm rate in hypothesis testing theory. These two schemes cannot help with the vulnerable cases where RSS from FBS is similar to that from the target BS as shown in Figure \ref{fig:4_4_3 Sequence Scheme compared with [42] and scheme with LF} and Figure \ref{fig:4_4_4 Sequence Scheme compared with [43]}.

\subsection{Potential Cost Discussion}
Although there have been considerations in the design of PSD Scheme not to largely affect the network efficiency, four main costs are inevitable in this PSD Scheme: 

(1) Memory space to store the table of sequence. Since the table of symbols is not temporary - selected symbols are generated for each handover, the table of symbols are known and stored in each LBS. Certain memory space will be used.


(2) Overload sequence information in Handover Preparation stage. Between the source BS and UE, \textit{Handover Request Ack} already includes a dedicated RACH preamble, access parameters, SIBs, etc. With long sequence of symbols to transmit, the regular transmit rate and efficiency can be impaired.

(3) Overload sequence information in Handover Execution stage between the target BS and UE. In response to UE synchronization request, the downlink transmission from the target BS already includes UL allocation, UL grant and timing advance. Symbols as headers are extra transmission overhead.

(4) Synchronization. The synchronization is realized by default. However, the PSD Scheme has higher requirements for synchronization than RSS-based schemes. It is sensitive to the inconsistency between desired symbols and received ones. Inaccurate synchronization may degrade detection effects.

\section{Conclusion}

In this paper, a PSD Scheme has been proposed for the detection of FBS based on the received signal sequence. Under the adversary scenario, the proposed scheme, which involves the analysis of table of symbols and selected sequence of symbols for verification, can hugely increase the detection accuracy and efficiency. Moreover, with the increase of table length comes the better detection effects under fixed sequence length. Finally, this PSD scheme is compared with several representative FBS detection schemes and overwhelmingly better performances are shown. In future work, we will consider the effects of potential costs on the PSD Scheme.

\bibliographystyle{IEEEtran}
\bibliography{references.bib}

\end{document}